\documentclass[fleqn,10pt]{wlscirep}
\usepackage[T1]{fontenc}  
\usepackage{multirow}
\usepackage{epsfig,epsf,psfrag} 
\usepackage{graphicx} 
\usepackage{subfigure}
\usepackage{amsmath}
\usepackage{hyperref}
\usepackage{pslatex}
\usepackage{epic,eepic} 
\usepackage{color,pstcol}
\usepackage{pstricks} 
\usepackage{fancyhdr}
\title{Role of helical edge modes in the chiral quantum anomalous Hall state} 
 \author[1]{Arjun Mani}
\author[1,*]{Colin Benjamin}

\affil[1]{School of Physical Sciences, National Institute of Science Education \& Research, HBNI, Jatni-752050,\ India}

\affil[*]{colin.nano@gmail.com}


\begin{abstract}
 Although indications are that a single chiral quantum anomalous Hall(QAH) edge mode might have been experimentally detected. There have been very many recent experiments which conjecture that a chiral QAH edge mode always materializes along with a pair of quasi-helical quantum spin Hall (QSH) edge modes. In this work we deal with a substantial 'What If?' question- in case the QSH edge modes, from which these QAH edge modes evolve, are not topologically-protected then the QAH edge modes wont be topologically-protected too and thus unfit for use in any applications. Further, as a corollary one can also ask if the topological-protection of QSH edge modes does not carry over during the evolution process to QAH edge modes then again our 'What if?' scenario becomes apparent. The 'how' of the resolution of this 'What if?' conundrum is the main objective of our work. We show in similar set-ups affected by disorder and inelastic scattering, transport via trivial QAH edge mode leads to quantization of Hall resistance and not that via topological QAH edge modes. This perhaps begs a substantial reinterpretation of those experiments which purported to find signatures of chiral(topological) QAH edge modes albeit in conjunction with quasi helical QSH edge modes.
 \end{abstract}
 
\begin{document}
\flushbottom
\maketitle

\begin{figure}
  \centering{ \includegraphics[width=0.7\textwidth]{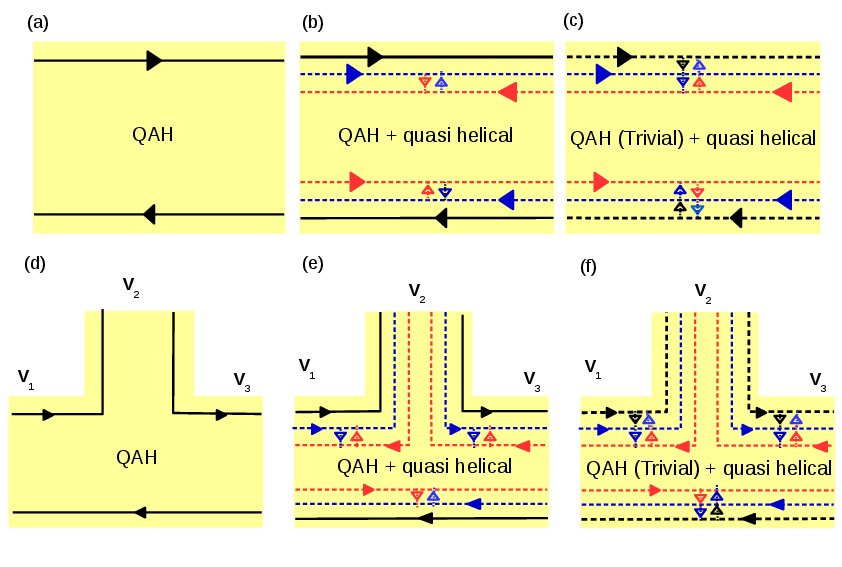}}
 \caption{Two and three terminal QAH bar: Topological chiral QAH edge mode is shown by solid black line, and quasi-helical QSH edge modes are shown by colored dashed line(red for spin up, blue for spin down). Trivial chiral QAH edge mode is shown by black dashed line. Intra edge back scattering is shown by arrows between the edge modes. Top panel: Two terminal QAH bar: (a) single topological QAH edge mode,  (b) single topological QAH edge mode with quasi-helical QSH edge modes and  (c) single trivial QAH edge mode with quasi-helical QSH edge modes, Bottom panel: Three terminal QAH bar: (d) single topological QAH edge mode, (e) single topological QAH edge mode with quasi-helical QSH edge modes and  (f) single trivial QAH edge mode with quasi-helical QSH edge modes.}
\end{figure}

\section*{Introduction}
Although, the experiment depicted in Ref.~\cite{cui} is most probably a detection of a single topological chiral quantum anomalous Hall(QAH) edge mode. There have been some other quite recent experiments\cite{kou,che,wang1} where it has been reported that QAH edge modes  occur in conjunction with quasi helical quantum spin Hall(QSH) edge modes\cite{wang}. The latter are prone to backscattering and are nothing but QSH edge modes which occur in a trivial insulator. These experiments which ``see'' QAH edge modes are in fact designed out of QSH edge mode setups in a topological insulator. By applying an extra Ferromagnetic layer or otherwise, an energy gap is sought to be created between the pair of helical edge modes in a QSH sample splitting these modes away from each other and suppressing one of these  leads to a single chiral QAH edge mode in a sample. However, contrary to expectation it is not just a chiral QAH mode which was seen in those experiments\cite{kou, che, wang1} but it always comes with the additional baggage of quasi-helical QSH edge modes in the trivial phase\cite{wang}. 

Helical QSH edge modes from which these chiral QAH edge modes evolve not only occur in topological insulators but they also do occur in a trivial insulator. Now applying a similar technique as before or attaching a  ferromagnetic layer to a trivial insulator, we can again make the trivial helical edge modes evolve into chiral QAH edge modes. But in the latter case, the chiral QAH edge mode so produced wont have a topological character and therefore this chiral QAH edge mode won't be protected against backscattering. Now this begs the question how can one be sure of the topological character of QAH edge modes.

Another question which can crop up is, does the topological nature of the QAH edge modes which evolve from helical QSH edge modes in a topological insulator survive the evolution. This ``evolution'' from helical QSH to chiral QAH edge mode as has been described in Refs.\cite{kou,che,wang1} is via addition of  magnetic impurities or a ferromagnetic layer. This may destroy their topological character since helical QSH edge modes are susceptible to spin flip scattering in presence of magnetic impurities. In this context  our work becomes relevant, since in those  QAH experiments what is quite evident is that the quantization of Hall resistance is  attributed to chiral topological QAH edge modes which exist in combination with quasi helical QSH edge modes. What our work reveals is that a chiral trivial QAH edge mode which exists in combination with quasi helical QSH edge modes gives the quantization of Hall resistance and not the chiral topological QAH edge mode when combined with trivial QSH edge modes. Thus  a shadow of doubt creeps up regarding the interpretation of those experiments\cite{wang}. 

We focus specifically on 4 and 6 terminal quantum anomalous Hall samples. We distinguish three cases one in which there is just a single chiral QAH edge mode which is topological in character (this hasn't been experimentally seen), the second wherein the chiral topological QAH edge mode exists alongwith  a pair of trivial QSH edge modes (this case is the supposed experimental result as in Refs.\cite{kou,che,wang}) and finally the case wherein a trivial QAH edge mode exists with a pair of trivial QSH edge modes (our 'What If?' scenario). Both the 4 terminal and 6 terminal samples are analyzed in three distinct regimes 1. where there is no disorder and inelastic scattering- the ideal case, 2. when there is disorder but no inelastic scattering and finally 3. when both disorder and inelastic scattering are present in the sample. The disorder we consider in our sample is restricted to terminal/contacts while inelastic scattering is present inside the sample and leads to the energy equilibration of the edge modes, see Refs.\cite{nikolajsen,Arjun,mani} for further details on energy equilibration as applied in different contexts in quantum Hall and quantum spin Hall samples. 

The rest of the paper is organized as follows- in section II we address the situation of two terminal and three terminal samples and distinguish between three cases of chiral topological QAH edge mode, chiral topological QAH edge mode with trivial  QSH edge modes and finally chiral trivial QAH edge mode with trivial  QSH edge modes. In section III we calculate the Hall resistance($R_H$),  two terminal local resistance $R_{2T}$ and finally the non-local resistance $R_{NL}$ for a four terminal sample with disorder and inelastic scattering and distinguish between the aforesaid three cases. In section IV we consider a six terminal sample and calculate the longitudinal resistance $R_L$ for afore-mentioned three different cases. We end our manuscript with a conclusion wherein we tabulate all the results of the 4 terminal and 6 terminal samples.

\section {Two terminal and three terminal QAH samples}
The Landauer-Buttiker formalism\cite{buti,datta}  relating currents with voltages in a multi terminal device has been extended to  QSH edge modes in Refs.~\cite{sanvito,chulkov} as well as  QAH samples in Ref.~\cite{wang}:
\begin{equation}
I_{i}=\sum_{j} ( G_{ji} V_{i} - G_{ij} V_{j})=\frac{e^{2}}{h} \sum_{j=1}^{N} ( T_{ji} V_{i} - T_{ij} V_{j}).
\end{equation}
In the above equation, $V_{i}$ is the voltage at i$^{th}$ probe/contact/terminal (we will be using the term probe or contact or terminal interchangeably for the same thing, i.e., a reservoir of electrons at some fixed potential) and $I_{i}$ is the current passing through the same terminal. $T_{ij}$ is the transmission probability from the $j^{th}$ to  $i^{th}$ probe and $G_{ij}$ is the conductance. $N$ denotes the number of terminals in the system.

\subsection{Chiral(topological) QAH edge mode (2 terminal)}
The case of a single chiral(topological) QAH edge mode is represented in Fig.~1(a). The relations between currents and voltages at the two terminals are derived from conductance matrix(2):
\begin{equation}
G_{ij} =\frac{e^{2}}{h} \left( \begin{array}{cc}
    1 & -1 \\
    -1  & 1  \\\end{array} \right),
\end{equation}

Choosing reference potential $V_{2}=0$, we derive the local (two probe) resistance  $R_{2T}^{QAH}=R_{12,12}=\frac{h}{e^{2}}$. 
\subsection{Chiral QAH edge mode with  trivial QSH edge modes (2 terminal)}
Here we have considered the general case, where the spin-flip scattering parameter $f_0$ between QAH chiral and trivial helical edge modes, will decide whether the QAH edge mode is topological (if $f_0=0$) or trivial (if $f_0\ne 0$). While $f$ defines the spin-flip scattering between the trivial helical edge modes. This situation of topological chiral plus trivial helical QSH is shown in Fig.~1(b) while case of trivial chiral QAH plus trivial helical QSH is shown in Fig.~1(c). The current-voltage relations can be easily derived from  conductance matrix below:
\begin{equation}
G_{ij} =\frac{e^{2}}{h} \left( \begin{array}{cc}
    (3-2(f+f_0)) & -(3-2(f+f_0)) \\
    -(3-2(f+f_0))  & (3-2(f+f_0))  \\\end{array} \right),
\end{equation}

Choosing reference potential $V_{2}=0$, we derive local (two probe) resistance  $R_{2T}^{Triv}=R_{12,12}=\frac{h}{e^{2}}\frac{1}{(3-2(f+f_0))}$. 
Setting parameter $f_0=0$ will give the two terminal resistance for topological QAH with trivial helical edge modes. The local resistance for topological QAH with trivial edge modes is $R_{2T}^{Topo}=R_{12,12}=\frac{h}{e^{2}}\frac{1}{(3-2f)}$. For maximal spin flip scattering $f=f_0=0.5$ we see that trivial $R_{2T}^{Triv}$ goes to the $R_{2T}^{QAH}$ which is equal to $2 R_{2T}^{Topo}$.

\subsection{Chiral (topological) QAH edge mode (3 terminal)}
The single chiral(topological) QAH edge mode case is represented in Fig. 1(d). The relations between currents and voltages at the various terminals are  derived from the conductance matrix below:
\begin{equation}
G_{ij} =\frac{e^{2}}{h} \left( \begin{array}{ccc}
    1 & 0 & -1 \\
   -1 & 1 & 0\\
   0  &  -1 & 1  \\\end{array} \right),
\end{equation}
\\
Since probe $2$ is the voltage probe, the current through probe $2$- $I_2$ is zero. We further choose reference potential $V_3=0$ which gives $V_{2}=V_{1}$. So, local (two terminal) resistance  $R_{2T}^{QAH}=R_{12,12}=\frac{h}{e^{2}}$. Because the QAH edge mode is chiral the voltage probe has no effect on it. 

\subsection{Chiral (topological) QAH edge mode with trivial QSH edge modes(3 terminal)}
The case of a single chiral(topological) QAH edge mode with a pair of quasi-helical QSH edge modes is shown in Fig. 1(e). The current-voltage relations can be derived from the conductance matrix below:
\begin{equation}
G_{ij} =\frac{e^{2}}{h} \left( \begin{array}{ccc}
    (3-2(f+f_0)) & -(1-(f+f_0))&  -(2-(f+f_0))\\
     -(2-(f+f_0)) & (3-2(f+f_0))& -(1-(f+f_0))\\
      -(1-(f+f_0))& -(2-(f+f_0))& (3-2(f+f_0))  \\\end{array} \right),
\end{equation}

Choosing reference potential $V_{3}=0$ and $I_2=0$ (as probe 2 is a voltage probe), we get $V_{2}=\frac{2-(f+f_0)}{(3-2(f+f_0))}V_{1}$. So, local (two terminal) resistance  $R_{2T}^{Triv}=R_{13,13}=\frac{h}{e^{2}}\frac{3-2(f+f_0)}{(7-9(f+f_0)+3(f+f_0)^2)}$. For topological($f_{0}=0$) case the local (two probe) resistance $R_{2T}^{Topo}=R_{13,13}=\frac{h}{e^{2}}\frac{3-2f}{(7-9f+3f^2)}$. The local (two probe) resistance is indeed affected by the voltage probe as seen from the two and three terminal cases. For maximal spin flip scattering $f=f_0=0.5$ we see that trivial $R_{2T}^{Triv}$ goes over to $R_{2T}^{QAH}$ which is equal to $\frac{13}{8}R_{2T}^{ Topo}$. Thus from looking at just 2T and 3T samples it is quite evident that the trivial QAH case crosses over to the the single chiral QAH case and not the topological QAH.\\

\section{Four Terminal quantum anomalous Hall bar}
The four terminal sample is shown in Fig.~2. We calculate the Hall resistance $R_{H}=R_{13,24}$, the local (two probe) resistance $R_{2T}=R_{13,13}$ and the non-local resistance $R_{NL}=R_{14,23}$ for various cases starting with just a single chiral(topological) QAH edge mode, then the chiral(topological) QAH edge mode with trivial quasi-helical QSH edge modes and finally the case of chiral(trivial) QAH edge mode with trivial quasi-helical QSH edge modes.

\subsection{Chiral topological QAH edge mode}

\subsubsection{Effect of  disorder}
Herein we consider two of the contacts (2,4) to be disordered, see Fig.~2(a).  Relations between the currents and voltages at the various terminals can be deduced from the conductance matrix, given below:
\begin{equation}
G_{ij} =\frac{e^{2} }{h} \left( \begin{array}{cccc}
    1  & 0 & -R_4 & -T_4 \\
    -T_2 & T_2 & 0 & 0\\ 
    -R_2  & -T_2 & 1 & 0 \\
     0  & 0 & -T_4 &  T_4 \\ \end{array} \right)
\end{equation}
Choosing reference potential $V_{3}=0$, further since 2 and 4 are voltage probes, $I_2=I_4=0$, we thus have $V_{2}=V_{1}$ and $V_3=V_4=0$. So, local (two probe) resistance  $R_{2T}^{QAH}=R_{13,13}=\frac{h}{e^{2}}$. Hall resistance- $R_H^{QAH}=R_{13,24}=\frac{V_2-V_4}{I_1}=\frac{h}{e^{2}}$. Disorder has no effect on the topological chiral QAH edge mode, the Hall resistance and local resistance remain the same as in the ideal(zero disorder) case. Finally, to calculate the non-local resistance $R_{NL}$ we consider $2,3$ as voltage probes and $1,4$ as current probes, we get $V_2=V_3$ which gives $R_{NL}=0$. Thus disorder has no effect on a single  chiral(topological) QAH edge mode.

\begin{figure}
 \centering {\includegraphics[width=.98 \textwidth]{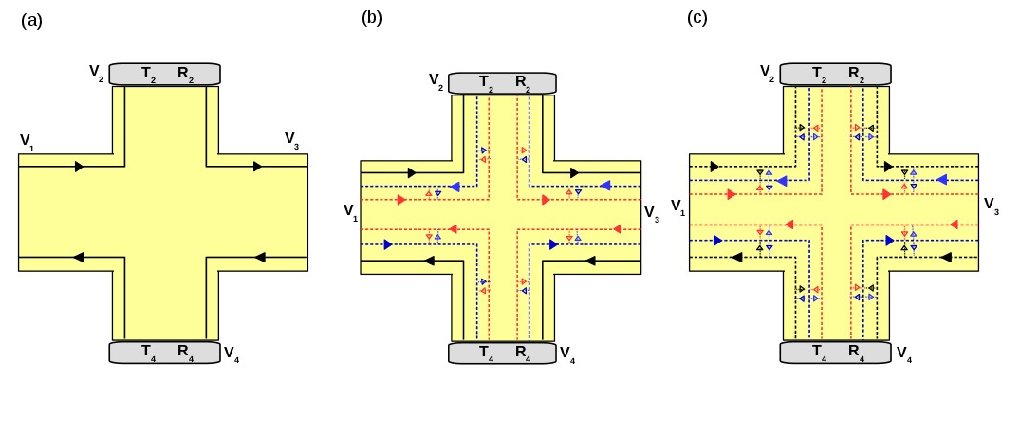}}
 \caption{Four terminal quantum anomalous Hall bar showing chiral QAH and quasi-helical QSH edge modes. (a) Topological QAH edge mode with disorder at probes $2$ and $4$ : $R_{2}, T_{2}$ and $R_4$, $T_4$ represent the reflection and transmission probability of edge modes from and into contact $2$ and $4$ respectively, (b) Topological QAH edge mode with quasi-helical edge modes: Disordered probes $2$ and $4$,  (c) Trivial QAH edge mode with  quasi-helical edge modes: Disordered probes $2$ and $4$.}
\end{figure} 

\subsubsection{Effect of disorder and inelastic scattering}
Similar to before, we consider two of the contacts ($2, 4$) are disordered, see Fig.~2(a). The electrons in-coming from probe $1$ with energy $\frac{e^2}{h}R_2V_1$ are equilibrated with the electrons coming from $2$ with energy $\frac{e^2}{h}T_2V_2$ to a new energy $\frac{e^2}{h}(R_2+T_2)V'_2=\frac{e^2}{h}V'_2$. Similarly electrons coming from probe $3$ are equilibrated with the electrons entering from probe $4$ to a new energy as shown below-
\begin{eqnarray} 
\frac{e^2}{h}R_2V_1+\frac{e^2}{h}T_2V_2=\frac{e^2}{h}V'_2,\qquad \frac{e^2}{h}V_1= \frac{e^2}{h}V'_1,\nonumber\\
\frac{e^2}{h}R_4V_3+\frac{e^2}{h}T_4V_4=\frac{e^2}{h}V'_4, \qquad  \frac{e^2}{h}V_3= \frac{e^2}{h}V'_3.
\end{eqnarray}
The currents and voltages at the contacts from 1 to 4 are related by the equations-
\begin{eqnarray}
I_1=\frac{e^2}{h}(V_1-V'_4)\nonumber\\
I_2=\frac{e^2}{h}T_2(V_2-V'_1)\nonumber\\
I_3=\frac{e^2}{h}(V_3-V'_2)\nonumber\\
I_4=\frac{e^2}{h}T_4(V_4-V'_3)
\end{eqnarray}

Choosing reference potential $V_{3}=0$ and $I_2=I_4=0$, since 2 and 4 are voltage probes, we thus derive $V_{2}=V_{1}$ and $V_3=V_4=0$. So, local (two probe) resistance  $R_{2T}^{QAH}=R_{13,13}=\frac{h}{e^{2}}$. The Hall resistance $R_H^{QAH}=R_{13,24}=\frac{V_2-V_4}{I_1}=\frac{h}{e^{2}}$. Similarly non-local resistance is derived as before-$R_{NL}^{QAH}=(V_2-V_3)/I_1=0$. So inelastic scattering too, like disorder at voltage probe has no effect on the a single chiral(topological) QAH edge mode.

\subsection{Chiral (topological) QAH edge mode with trivial QSH edge modes}

\subsubsection{Effect of disorder} Herein,  as before we consider two of the contacts $2$ and $4$ to be disordered, see Fig.~2(b). The relations between currents and voltages at the various terminals can be obtained from the conductance matrix below:
\begin{equation}
G_{ij} =\frac{e^{2} }{h} \left( \begin{array}{cccc}
    T^{11}  & -T^{12} & -T^{13} & -T^{14} \\
    -T^{21}  & -T^{22} & -T^{23} & T^{24} \\ 
    T^{31}  & T^{32} & T^{33} & T^{34} \\
     T^{41}  & T^{42} & T^{43} & T^{44} \\ \end{array} \right)
\end{equation}
where 
\begin{eqnarray} 
 T^{11}&=&(3 - 2 f  - a_1 R_2^2 (1 - f )/(a) - R_4^2 (1 - f) a_1/(c))\nonumber\\
T^{12}&=&(1 - f) T_2/(1-R_2f)\nonumber\\
T^{13}&=&((1 - f)^2 R_2/a + ((1 - f)^2 + (1 - f)^2) R_4 + R_4^3 a_1^2/c)\nonumber\\
T^{14}&=&((-2 + f + f R_4) T_4)/(-1 + f R_4)\nonumber\\
T^{21}&=&((-2 + f + f R_2) T_2)/(-1 + f R_2)\nonumber\\
T^{22}&=&T_2 (3 - 2 f T_2/(1 - f R_2))\nonumber\\
T^{23}&=&(1 - f) T_2/(1-R_2f), \qquad\qquad \qquad T^{24}=0,
\label{4t-trans}
\end{eqnarray}
with $a = 1 - R_2^2 f^2, c = 1 - R_4^2 f^2,  a_1 =  f (1 - f)$. By interchanging $R_2$ and $R_4$ in the above expressions for $T^{11}$, $T^{12},.., T^{23}$ rest of the transmission probabilities $T^{31}$ to $T^{44}$ can be deduced. The transmission probabilities are calculated in this way- say $T^{23}$, the transmission probability of electron from terminal $3$ to $2$ can be explained as the sum of paths available from $3$ to $2$ for one chiral topological edge mode and one pair of trivial helical edge modes. An electron in the topological edge mode coming out of probe $3$ has probability zero to reach probe $2$. But an electron in the trivial helical edge mode has finite probability to reach probe $2$ from $3$. An electron coming out of probe $3$ can reach probe $2$ with probability $T_2(1-f)$, but that is just one path, it can also reach $2$ with probability $fR_2T_2(1-f)$  following a second path due to spin flip scattering, similarly a third path is $f^2R_2^2T_2(1-f)$. Thus, we can form an infinite number of paths from probe $3$ to $2$, these can be summed to get the total transmission probability as $T^{23}=\frac{T_2(1-f)}{(1-R_2f)}$. Similarly, the other transmission probabilities in Eq.~\ref{4t-trans} are obtained.
 
Choosing reference potential $V_{3}=0$, and since $2$ and $4$ are voltage probes, we derive the local (two probe) resistance in absence of disorder ($R_2=R_4=0$) as $R_{2T}^{Topo}=R_{13,13}=\frac{3-2f}{5-6f+2f^2}$. The Hall resistance- $R_H^{Topo}=R_{13,24}=\frac{h}{e^{2}}\frac{1}{(5 + 2 f^2 -6f )}$. Similarly, as before non-local resistance is deduced as $R_{NL}^{Topo}=\frac{(2-f)(1-f)(3-2f)}{(5-6f+2f^2)(7-9f+3f^2)}$. For general case (i.e., with disorder) the expressions for $R_H$, $R_{2T}$ and $R_{NL}$ are too large to be reproduced here, so we will analyze them via plots, see Figs.~3(a-d).\\

\begin{figure}
 \centering { \includegraphics[width=.6\textwidth]{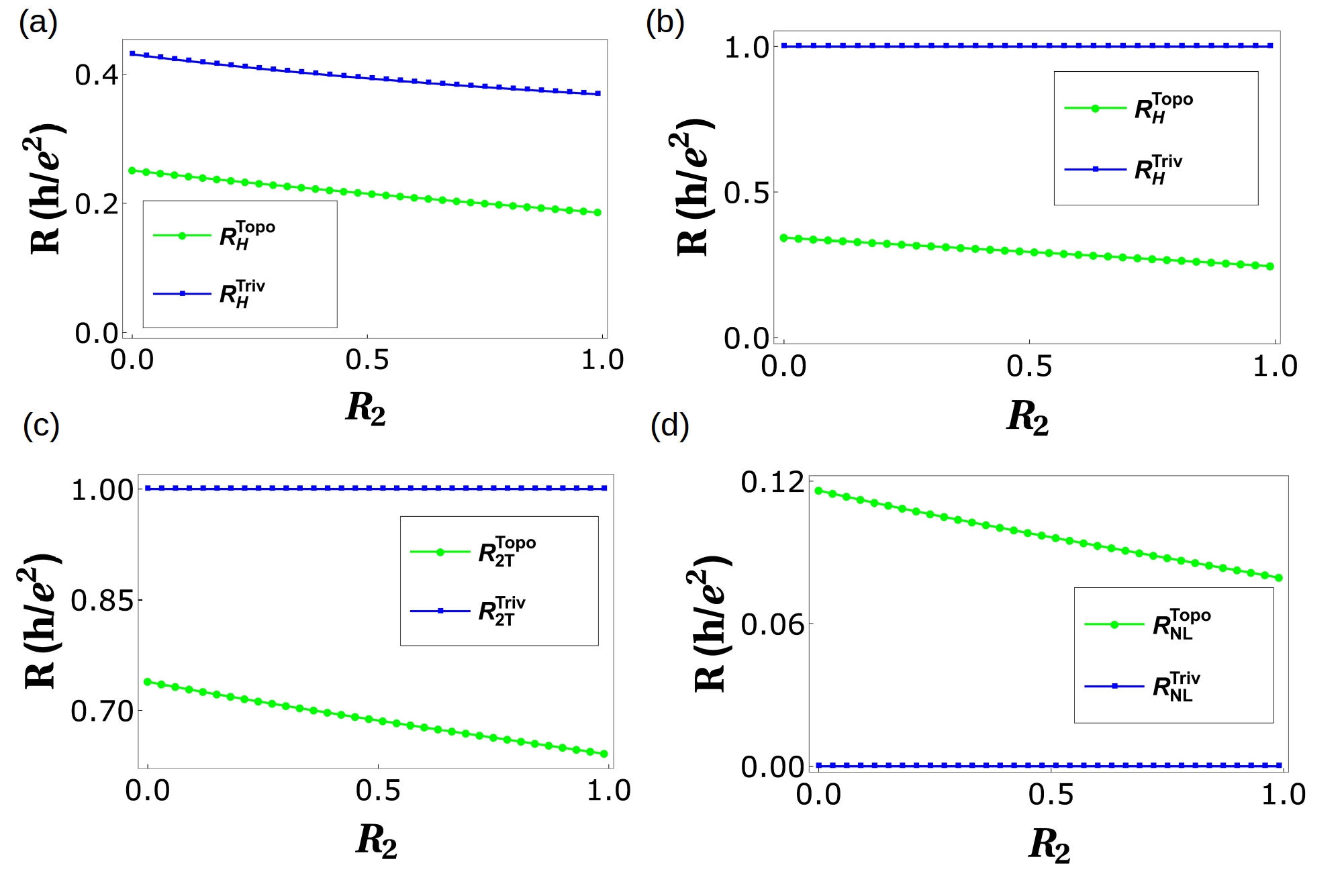}}
\caption{$R_{NL}$, $R_{2T}$ and  $R_{H}$ vs. Disorder. (a) Hall resistance  vs. Disorder $R_2$ with $R_4=0.5$ and spin-flip probability $f=0.3$ and $f_0=0.3$, (b)  Hall resistance  vs. disorder $R_2$ with $R_4=0.5$ and spin-flip probability $f=0.5$ and $f_0=0.5$. (c) Two-terminal resistance  vs. Disorder $R_2$ with $R_4=0.5$ and spin-flip probability$f=0.5$ and $f_0=0.5$, (d) Non-local resistance  vs. Disorder $R_2$ with $R_4=0.5$ and spin-flip probability $f=0.5$ and $f_0=0.5$.  }
\end{figure} 

\subsubsection{Effect of disorder and inelastic scattering}
Herein we consider the effect of both disorder and inelastic scattering on topological QAH edge modes as shown in Fig.~4(a). Here the inelastic scattering is shown by starry blobs as in Fig.~4(a). As the QAH edge mode is topological, it will not equilibrate its energy with trivial helical edge modes. Thus, topological chiral edge modes equilibrate only between themselves, these equilibrate to energy $V''_i$ where $ i $ is the contact index from 1 to 4. The trivial helical edge modes equilibrate with  other trivial helical edge modes and these equilibrate their energy to $V'_i$. The contacts $2$ and $4$ are disordered as in the previous case. The currents and voltages at the contacts from 1 to 4 are related by the equations-
\begin{eqnarray}
I_1&=&(3-2f)V_1-V''_4-(1-f)(V'_1+V'_4),\nonumber\\
I_2&=&(3T_2-\frac{2T_2^2f}{(1-R_2f)})V_2-T_2V''_1-\frac{T_2(1-f)}{(1-R_2f)}(V'_1+V'_2),\nonumber\\
I_3&=&(3-2f)V_3-V''_2-(1-f)(V'_2+V'_3),\nonumber\\
I_4&=&(3T_4-\frac{2T_4^2f}{(1-R_4f)})V_4-T_4V''_3-\frac{T_4(1-f)}{(1-R_2f)}(V'_3+V'_4),\nonumber\\
\end{eqnarray}
where the potentials $V'_i$ and $V''_i$ are related to $V_i$ by-
\begin{eqnarray}
V''_1=V_1 ,\qquad R_2V''_1+T_2V_2=V''_2,\nonumber\\
V''_3=V_3 ,\qquad R_2V''_3+T_2V_4=V''_4,
\end{eqnarray}
and 
{
\begin{eqnarray}
&(1-f)V_1+\frac{T_2(1-f)}{(1-R_2f)}V_2+\frac{R_2(1-f)^2}{a}V'_2=((1-f)+\frac{T_2(1-f)}{(1-R_2f)}+\frac{R_2(1-f)^2}{a})V'_1,\nonumber\\
&(1-f)V_3+\frac{T_2(1-f)}{(1-R_2f)}V_2+\frac{R_2(1-f)^2}{a}V'_1=((1-f)+\frac{T_2(1-f)}{(1-R_2f)}+\frac{R_2(1-f)^2}{a})V'_2,\nonumber\\
&(1-f)V_3+\frac{T_4(1-f)}{(1-R_4f)}V_4+\frac{R_4(1-f)^2}{c}V'_4=((1-f)+\frac{T_4(1-f)}{(1-R_4f)}+\frac{R_4(1-f)^2}{c})V'_3,\nonumber\\
&(1-f)V_1+\frac{T_4(1-f)}{(1-R_4f)}V_4+\frac{R_4(1-f)^2}{c}V'_3=((1-f)+\frac{T_4(1-f)}{(1-R_4f)}+\frac{R_4(1-f)^2}{c})V'_4,\nonumber\\
\end{eqnarray}

with $a = 1 - R_2^2 f^2, c = 1 - R_4^2 f^2$. Choosing reference potential $V_{3}=0$, and the contact $2$ and $4$ to be voltage probe as before, i.e., $I_2=I_4=0$, we derive local (two probe) resistance in absence of disorder as $R_{2T}^{Topo}=R_{13,13}=\frac{4-2f}{5-4f+f^2}$ (for $R_2=R_4=0$). The Hall resistance $R_H^{Topo}=R_{13,24}=\frac{h}{e^{2}}\frac{2}{(5 -4f+f^2)}$. For the general case (including disorder) the expressions for $R_H$, $R_{2T}$ and $R_{NL}$ are again large, so we will analyze them via plots, see Figs.~5(a-d). 

\begin{figure}
 { \includegraphics[width=.9\textwidth]{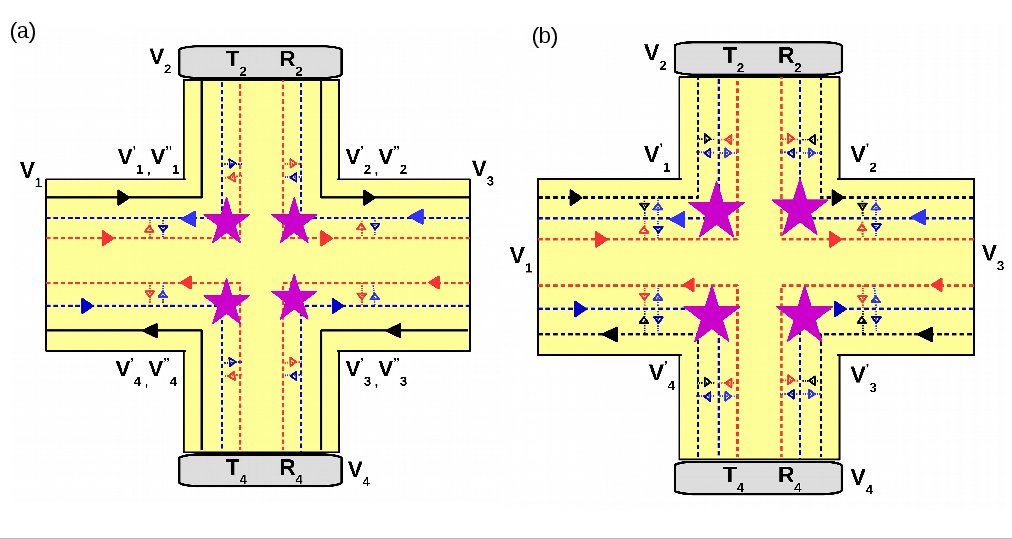}}
 \caption{Four terminal QAH sample with disorder and inelastic scattering. (a) Chiral (topological) QAH edge mode with quasi-helical QSH edge modes in presence of inelastic scattering and disorder, (b) Chiral (trivial) QAH edge mode with quasi-helical QSH edge modes in presence of inelastic scattering and disorder. }
\end{figure} 

\subsection{Chiral (trivial) QAH edge mode  with trivial  QSH edge modes}

\subsubsection{Effect of  disorder } Herein as before we consider two of the contacts $2$ and $4$ to be disordered, see Fig.~2(c). The current-voltage relations are derived from the conductance matrix below:
\begin{equation}
G_{ij} =\frac{e^{2} }{h} \left( \begin{array}{cccc}
    T^{11}  & -T^{12} & -T^{13} & -T^{14} \\
    -T^{21}  & T^{22} & -T^{23} & -T^{24} \\ 
    -T^{31}  & -T^{32} & T^{33} & -T^{34} \\
     -T^{41}  & -T^{42} & -T^{43} & T^{44} \\ \end{array} \right),
\end{equation}
where 
\begin{eqnarray}
 T^{11}&=&(3 - 2 (f + f_0) - a_1 R_2^2 (1 - f - f_0)/(a) - R_4^2 (1 - f - f_0) a_1/(c)),\nonumber\\
T^{12}&=&((1 - f - f_0) T_2/a + (1 - f - f_0) T_2 (f + f_0) R_2/a),\nonumber\\
T^{13}&=&((1 - f - f_0)^2 R_2/a + ((1 - f)^2 + (1 - f_0)^2) R_4 + R_4^3 a_1^2/c),\nonumber\\
T^{14}&=&(T_4 (2 - f - f_0) + (f + f_0) T_4 R_4^2 a_1/c + T_4 R_4 a_1/c),\nonumber\\
T^{21}&=&((2 - f - f_0) T_2 + T_2 R_2 a_1/a + T_2 R_2^2 a_1 (f + f_0)/a),\nonumber\\
T^{22}&=&(3 T_2 - T_2^2 (f + f_0)/a - T_2 ^2 R_2 (f + f_0)^2/a - T_2^2 (f + f_0)/a - T_2^2 R_2 (f^2 + f_0^2)/a),\nonumber\\
T^{23}&=&((1 - f - f_0) T_2 (R_2 (f + f_0)/a + 1/a)),\nonumber\\
T^{24}&=&0.
\end{eqnarray}
with $a = 1 - R_2^2 (f^2 + f_0^2), c = 1 - R_4^2 (f^2 + f_0^2),  a_1 =  f (1 - f) + f_0 (1 - f_0)$. By interchanging $R_2$ and $R_4$ in the above equation(15) rest of the transmission probabilities $T^{31}$ to $T^{44}$ can be deduced. The transmission probabilities can be  explained as before. $T^{23}$, the transmission probability of electron from terminal $3$ to $2$ can be explained as the sum of probabilities from $3$ to $2$ for all the edge modes over all possible paths. An electron coming out of probe $3$ at upper edge can reach probe $2$ with probability $T_2(1-f-f_0)$, but that is just one path, it can also reach $2$ with probability $(f+f_0)R_2T_2(1-f-f_0)$  following a second path due to spin flip scattering, similarly probability for a third path is $(f^2+f_0^2)R_2^2T_2(1-f-f_0)$. These first, third, fifth.. paths form an infinite series with total transmission probability $\frac{T_2(1-f-f_0)}{a}$ and second, fourth, sixth... paths form a infinite series with total transmission probability $\frac{(f+f_0)R_2T_2(1-f-f_0)}{a}$. So, the total transmission probability is sum of the two and is written as  $T^{23}$ as in Eq.~(15).
 
Choosing reference potential $V_{3}=0$, and $I_2= I_4=0$ (as $2$ and $4$ are voltage probes) we derive local (two probe) resistance in absence of any disorder as $R_{2T}^{Triv}=R_{13,13}=\frac{3-2(f+f_0)}{5-6(f+f_0)+2(f+f_0)^2}$. The Hall resistance- $R_H^{Triv}=R_{13,24}=\frac{h}{e^{2}}\frac{1}{(5 + 2 f^2 + 2 (-3 + f_0) f_0 + f (-6 + 4 f_0))}$ again for zero disorder. Similarly, as before  the non-local resistance is deduced as $R_{NL}^{Triv}=\frac{(2-f-f_0)(1-f-f_0)(3-2f-2f_0)}{(5-6(f+f_0)+2(f+f_0)^2)(7-9(f+f_0)+3(f+f_0)^2)}$ for zero disorder. For, general case the expressions for $R_H$, $R_{2T}$ and $R_{NL}$ are again too large to be reproduced here, so we will examine them via plots, see Figs.~3(a-d).

\subsubsection{ Effect of  disorder and inelastic scattering}
Herein, we consider the trivial QAH edge modes with both disorder and inelastic scattering as shown in Fig.~4(b). Here the QAH chiral as well as helical both edge modes are in the trivial phase, i.e., they  are all prone to intra edge back scattering due to spin-flips. All the edge modes interact among themselves leading to their energies being equilibrated to the potential $V'_i$ ($`i'$ is from 1 to 4). The contacts $2$ and $4$ are disordered as in the previous case. The currents and voltages at the contacts from 1 to 4 are related by the equations-
{
\begin{eqnarray}
I_1&=&[3-2(f+f_0)]V_1-[1-(f+f_0)]V'_1-[2-(f+f_0)]V'_4 ,\nonumber\\
I_2&=&T_2(3-\frac{(2(f+f_0+(f^2 + f f_0 + f_0^2) R_2) T_2)}{(1 - R_2^2(f^2 + f_0^2) ))}V_2-\frac{[1-(f+f_0)][1+R_2(f+f_0)]T_2}{[1-R_2^2(f^2+f_0^2)]}V'_2-\nonumber\\&&\frac{[2 - (f + f_0) + ((1 - f) f + (1 - f_0) f_0) R_2 - (f - f_0)^2 R_2^2) T_2}{[1-R_2^2(f^2+f_0^2)]}V'_1 ,\nonumber\\
I_3&=&[3-2(f+f_0)]V_3-[1-(f+f_0)]V'_3-[2-(f+f_0)]V'_2,\nonumber\\
I_4&=&T_2(3-\frac{(2(f+f_0+(f^2 + f f_0 + f_0^2) R_4) T_4)}{(1 - R_4^2(f^2 + f_0^2) ))}V_4-\frac{[1-(f+f_0)][1+R_4(f+f_0)]T_4}{[1-R_4^2(f^2+f_0^2)]}V'_4-\nonumber\\&&\frac{[2 - (f + f_0) + ((1 - f) f + (1 - f_0) f_0) R_4 - (f - f_0)^2 R_4^2) T_2}{[1-R_4^2(f^2+f_0^2)]}V'_3\nonumber.\\
\end{eqnarray}
The relations between potentials $V'_i$'s and $V_i$ are written in the appendix.


Choosing reference potential $V_{3}=0$, and as before $I_2=I_4=0$ (these are voltage probes), we derive the local (two probe) resistance  $R_{2T}^{Triv}=R_{13,13}$  and the Hall resistance $R_H^{Triv}=R_{13,24}$. Similarly, as before the non-local resistance is deduced as $R_{NL}^{Triv}=R_{14,23}$. The expressions for $R_H$, $R_{2T}$ and $R_{NL}$ are large, so again we will analyze them via plots as in Figs.~5(a-d). 
\begin{figure}
  \centering    { \includegraphics[width=.6\textwidth]{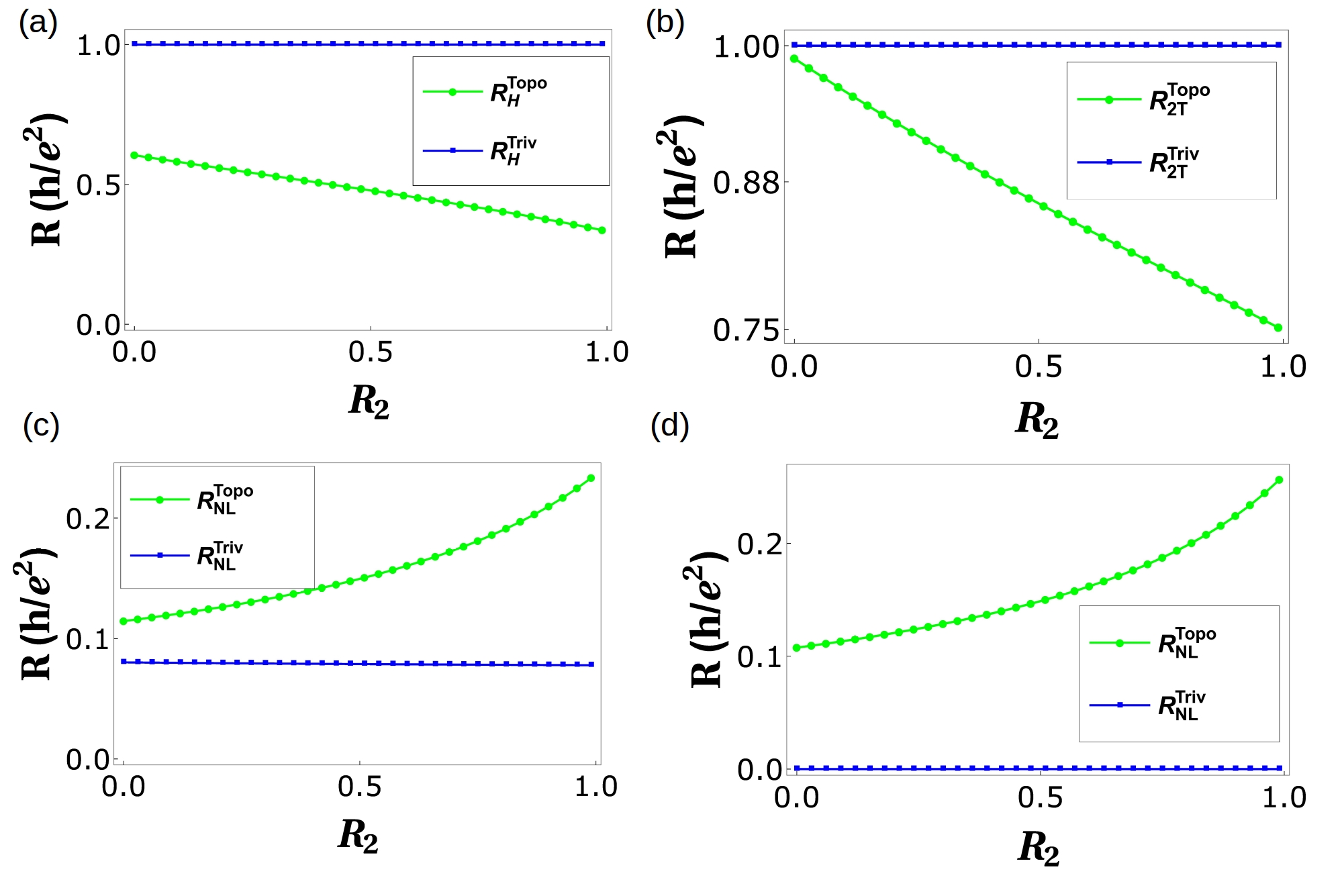}}
\caption{$ R_{H}$, $R_{2T}$ and $R_{NL}$  under the effect of  inelastic scattering. (a) Hall resistance  vs. Disorder $R_{2}$ with parameters $R_4=0.5$ and spin-flip probability $f=f_0=0.5$, (b) Two terminal resistance  vs. Disorder $R_{2}$  with parameters $R_4=0.5$ and spin-flip probability $f=0.5$, $f_0=0.5$, (c) Non-local resistance  vs. Disorder $R_{2}$  parameters $R_4=0.5$ and spin-flip probability $f=0.3$, $f_0=0.3$, (d) Non-local resistance  vs. Disorder $R_{2}$ parameters $R_4=0.5$ and spin-flip probability $f=0.5$, $f_0=0.5$.}
\end{figure} 
{
\begin{center}
\begin{table}
\caption{Comparison of  chiral(topological) QAH edge modes, chiral(topological) QAH  edge mode with trivial QSH edge modes and chiral(trivial) QAH  edge mode with trivial QSH edge modes}
\begin{tabular}{ |c|c|c|c|p{3.5cm}|}
 \hline
&&QAH(topological)&QAH(topological)+Trivial QSH&QAH(trivial)+Trivial QSH \\ 
\hline
\multirow{6}{*}{$R_H$}&Ideal (zero disorder) &Quantized $\frac{e^2}{h}$&Not quantized & Not quantized \\
&&$R_H(\uparrow)=-R_H(\downarrow)$&$R_H(\uparrow)=-R_H(\downarrow)$&$R_H(\uparrow)=-R_H(\downarrow)$\\
\cline{2-5}
&Finite disorder & Quantized $\frac{e^2}{h}$&Not quantized & Not quantized\\
&&$R_H(\uparrow)=-R_H(\downarrow)$&$R_H(\uparrow)=-R_H(\downarrow)$&$R_H(\uparrow)=-R_H(\downarrow)$\\
\cline{2-5}
&Disorder+Inelastic scattering & Quantized $\frac{e^2}{h}$& Not quantized (Fig.~5(a))&Quantized (Fig.~5(a))\\
&&$R_H(\uparrow)=-R_H(\downarrow)$&$R_H(\uparrow)\ne-R_H(\downarrow)$&$R_H(\uparrow)\ne-R_H(\downarrow)$\\
\hline
\hline
\multirow{6}{*}{$R_{2T}$}&Ideal (zero disorder) &Quantized $\frac{e^2}{h}$&Not quantized & Not quantized \\
&&$R_{2T}(\uparrow)=R_{2T}(\downarrow)$&$R_{2T}(\uparrow)=R_{2T}(\downarrow)$&$R_{2T}(\uparrow)=R_{2T}(\downarrow)$\\
\cline{2-5}
&Finite disorder &Quantized $\frac{e^2}{h}$&Not quantized (Fig.~3(b))& Not quantized (Fig.~3(b))\\
&&$R_{2T}(\uparrow)=R_{2T}(\downarrow)$&$R_{2T}(\uparrow)=R_{2T}(\downarrow)$&$R_{2T}(\uparrow)=R_{2T}(\downarrow)$\\
\cline{2-5}
&Disorder+Inelastic scattering&Quantized $\frac{e^2}{h}$&Not quantized (Fig.~5(b))& Quantized (Fig.~5(b))\\
&&$R_{2T}(\uparrow)= R_{2T}(\downarrow)$&$R_{2T}(\uparrow)\ne R_{2T}(\downarrow)$&$R_{2T}(\uparrow)\ne R_{2T}(\downarrow)$\\
\hline
\hline
\multirow{6}{*}{$R_{NL}$}&Ideal (zero disorder) & 0& Finite & Finite \\
&&$R_{NL}(\uparrow)=R_{NL}(\downarrow)$&$R_{NL}(\uparrow)=R_{NL}(\downarrow)$&$R_{NL}(\uparrow)=R_{NL}(\downarrow)$\\
\cline{2-5}
&Finite disorder &0&Finite (Fig.~3(c))&Finite (Fig.~3(c))\\
&&$R_{NL}(\uparrow)=R_{NL}(\downarrow)$&$R_{NL}(\uparrow)=R_{NL}(\downarrow)$&$R_{NL}(\uparrow)=R_{NL}(\downarrow)$\\
\cline{2-5}
&Disorder+Inelastic scattering&0&Finite (Fig.~5(c, d))& 0 (Fig.~5( d))\\
&&$R_{NL}(\uparrow)= R_{NL}(\downarrow)$&$R_{NL}(\uparrow)\ne R_{NL}(\downarrow)$&$R_{NL}(\uparrow)= R_{NL}(\downarrow)$\\
\hline
\end{tabular}
\end{table}
\end{center}
}

In table I we tabulate the results obtained so far. One important thing left out of our discussion so far has been the role of spin in QAH edge mode. A single chiral (topological) QAH should satisfy the following symmetry relations for $R_H(\uparrow)=-R_H(\downarrow)$ and $R_{NL}=0$. We  see that $R_{NL}(\uparrow)\neq R_{NL}(\downarrow)$ for  topological QAH(with quasi helical QSH) while this isn't case for trivial QAH(with quasi helical QSH) edge modes, see Fig.~\ref{nlspin-fig}. Importantly while in case of trivial QAH edge mode $R^{Triv}_{NL}(\uparrow)=R^{Triv}_{NL}(\downarrow)=0$, for the case of topological QAH edge mode $R^{Topo}_{NL}(\uparrow)\neq R^{Topo}_{NL}(\downarrow)\neq 0$.
In fact we see that $R^{Triv}_{NL}$ for the case of disorder and inelastic scattering approaches zero similar to a single chiral QAH edge mode, while for the $R^{Topo}_{NL}$ this doesn't, again leading to a contradiction with the way the experiments of Ref.~\cite{kou} have been interpreted as in Ref.~\cite{wang}. 
\begin{figure}
  \centering    { \includegraphics[width=.6\textwidth]{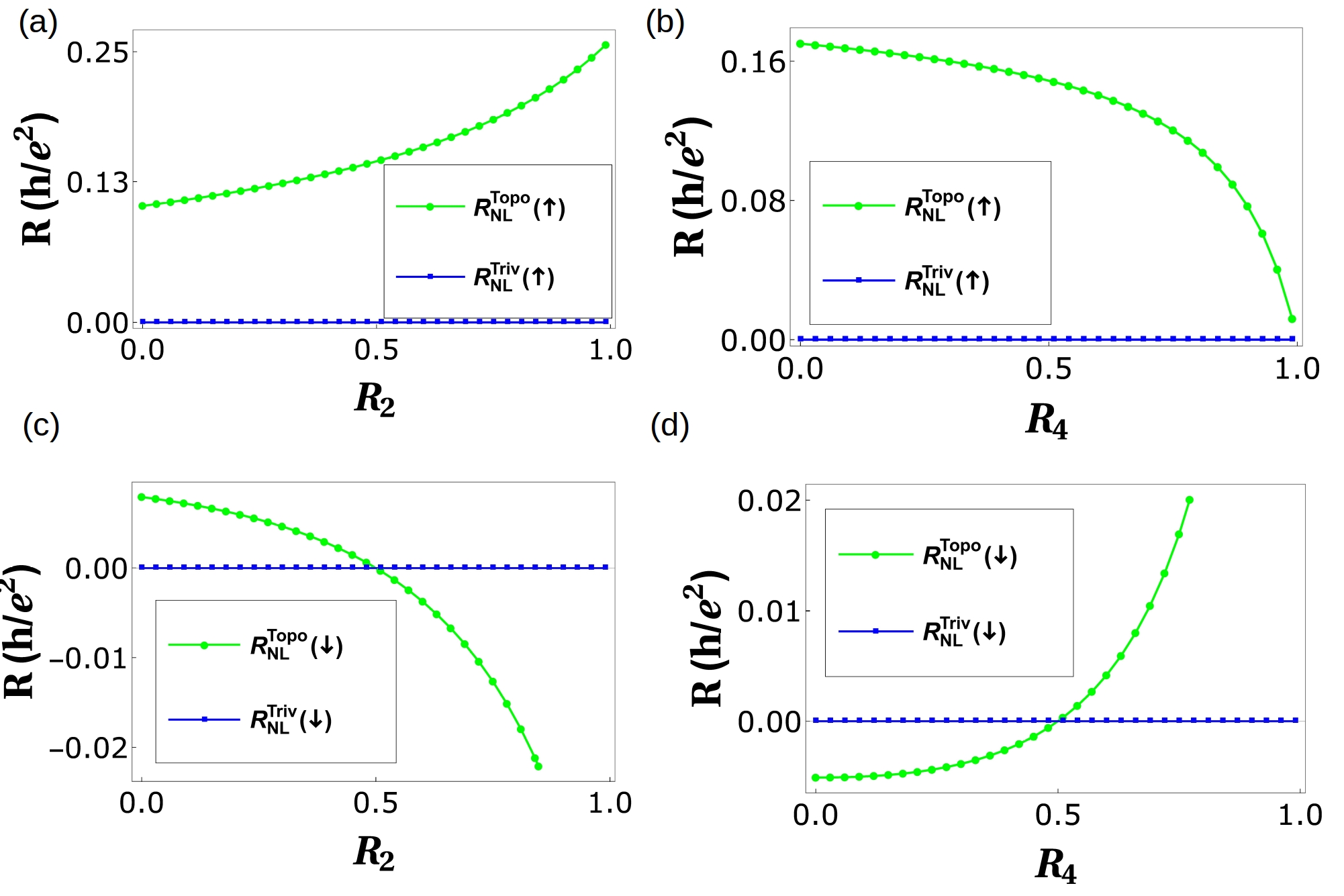}}
\caption{Non-local resistances ($R_{NL}(\uparrow)$) for magnetization direction ($\uparrow$) and ($R_{NL}(\downarrow)$) for magnetization direction ($\downarrow$) in presence of both disorder as well as inelastic scattering. (a) $R_{NL}(\uparrow)$ for chiral(topological) QAH edge mode with quasi-helical QSH edge modes and chiral (trivial) QAH edge mode with quasi-helical QSH edge modes for parameters $R_{4}= 0.5, f = f_{0} = 0.5$, (b) $R_{NL}(\uparrow)$ for chiral(topological) QAH edge mode with quasi-helical QSH edge modes and chiral (trivial) QAH edge mode with quasi-helical QSH edge modes for parameters $R_{2} = 0.5, f = f_{0} = 0.5$. (c) $R_{NL}(\downarrow)$ for chiral(trivial) QAH edge mode with quasi-helical QSH edge modes and chiral(trivial) QAH edge mode with quasi-helical QSH edge modes for parameters $R_{4}=0.5, f = f_{0} = 0.5$, (d) $R_{NL}(\downarrow)$ for chiral(trivial) QAH edge mode with quasi-helical QSH edge modes and chiral(trivial) QAH edge modes with quasi-helical edge modes for parameters $R_{2} = 0.5, f = f_{0} = 0.5$. }
\label{nlspin-fig}
\end{figure} 

\section{Six terminal quantum anomalous Hall bar}

In this section we analyze a six terminal QAH bar, we especially focus on the longitudinal resistance $R_{L}$. For a single chiral(topological) QAH edge mode $R_{L}=0$, but the experiments\cite{kou,che,wang1} revealed a  finite longitudinal resistance. This result prompted the interpretation of the experiments \cite{kou,che,wang1} as seeing not just a chiral(topological) QAH edge mode but in addition also a pair of quasi-helical QSH edge modes\cite{wang}. Since a non zero $R_L$ is the hallmark of helical QSH edge modes. Here we probe further by comparing as in sections before the three cases and try to find out if a topological QAH edge mode or a trivial QAH edge mode occurring with quasi-helical edge modes results in a non-zero $R_L$.  

\subsection{Chiral(topological) QAH edge mode}

\subsubsection{Effect of  disorder} Herein we consider two of the contacts 1 and 4 to be disordered. The  relations between the currents and voltages at the various terminals  can be derived from the conductance matrix below:
\begin{equation}
G_{ij} =\frac{e^{2} }{h} \left( \begin{array}{cccccc}
    T_1  & 0 & 0 & 0&0&-T_1 \\
    -T_1 & 1 & 0 & 0&0&-R_1\\ 
   0  & -1 & 1 & 0&0&0 \\
     0  & 0 & -T_4 &  T_4&0&0 \\
      0  & 0 & -R_4 &  -T_4&1&0 \\
      0  & 0 & 0&  0&-1&1 \\
      \end{array} \right)
\end{equation}
Choosing reference potential $V_{4}=0$ and $I_2=I_3=I_5=I_6=0$ (as these are voltage probes), we get $V_{3}=V_{2}=\frac{T_1V_1}{1-R_1R_4}$ and $V_3=V_4=\frac{T_1R_4V_1}{1-R_1R_4}$. So the longitudinal resistance $R_L^{QAH}=(V_2-V_3)/I_1=0$. So disorder has no effect on the longitudinal resistance for a single chiral QAH edge mode.

\subsubsection{Effect of disorder and inelastic scattering}
Herein we consider two of the contacts ($1, 4$) to be disordered. The electrons incoming from probe $6$ with energy $\frac{e^2}{h}R_1V'_6$ are equilibrated with the electrons incoming from probe $1$ with energy $\frac{e^2}{h}T_1V_1$ to a new energy $\frac{e^2}{h}(R_1+T_1)V'_1=\frac{e^2}{h}V'_1$. Similarly electrons coming from probe $3$ are equilibrated to the electrons coming from probe $4$ to a new energy as shown below-
\begin{eqnarray} 
\frac{e^2}{h}R_1V'_6+\frac{e^2}{h}T_1V_1&=\frac{e^2}{h}V'_1,\qquad \frac{e^2}{h}V_6= \frac{e^2}{h}V'_6,\nonumber\\
\frac{e^2}{h}R_4V'_3+\frac{e^2}{h}T_4V_4&=\frac{e^2}{h}V'_4, \qquad  \frac{e^2}{h}V_3= \frac{e^2}{h}V'_3,\nonumber\\
\frac{e^2}{h}V'_5&=\frac{e^2}{h}V_5, \qquad  \frac{e^2}{h}V'_2=\frac{e^2}{h}V_2.
\end{eqnarray}
The currents and voltages at the contacts from 1 to 4 are related by the equations-
\begin{eqnarray}
I_1=\frac{e^2}{h}T_1(V_1-V'_6),\qquad I_2=\frac{e^2}{h}(V_2-V'_1),\nonumber\\
I_3=\frac{e^2}{h}(V_3-V'_2),\qquad I_4=\frac{e^2}{h}T_4(V_4-V'_3),\nonumber\\
I_5=\frac{e^2}{h}(V_5-V'_4),\qquad I_6=\frac{e^2}{h}(V_6-V'_5).
\end{eqnarray}

Choosing the reference potential $V_{4}=0$ and $I_2=I_4=I_5=I_6=0$ (as these are voltage probes), we thus derive $V_{3}=V_{2}=V'_1=V'_2$ which gives the longitudinal resistance  $R_{L}^{QAH}=R_{14,14}=0$. 

\subsection{Chiral(topological) QAH edge mode with trivial  QSH edge modes}
\subsubsection{ Effect of disorder} As before we consider two of the contacts $1$ and $4$ to be disordered, see Fig.~7(a). The relations between currents and voltages at the various terminals can be derived from the conductance matrix below:
\begin{equation}
G_{ij} =\frac{e^{2} M}{h} \left( \begin{array}{cccccc}
    T^{11}  & -T^{12} & -T^{13} &-T^{14} &-T^{15} &-T^{16}\\
    -T^{21}  & T^{22} & -T^{23} & -T^{24} & -T^{25} & T^{26}\\ 
    -T^{31}  &- T^{32} & T^{33} & -T^{34} &-T^{35}  &- T^{36}  \\
    - T^{41}  & -T^{42} & -T^{43} & T^{44} & - T^{45}  & -T^{46} \\
     - T^{51}  & -T^{52} & -T^{53} & T^{54} & T^{55}  & -T^{56} \\
      - T^{61}  & -T^{62} & -T^{63} & -T^{64}  &- T^{65}  & T^{66}
    \\ \end{array} \right)
\end{equation}
where,
\begin{eqnarray}
 T^{11}&=&T_1(3-f(2+R_1))/(1-fR_1),\nonumber\\
T^{12}&=&T^{61}=T_1(1-f)/(1-fR_1),\nonumber\\
T^{13}&=&0,\qquad T^{14}=0,\qquad T^{15}=0,\nonumber\\
T^{16}&=&T^{21}=T_1(2-f-fR_1)/(1-fR_1),\nonumber\\
T^{22}&=&T^{66}=(3-2f)+(1-f^2)fR_1^2/(1-f^2R_1^2),\nonumber\\
T^{23}&=&(1-f),\qquad T^{24}=0,\qquad T^{25}=0,\nonumber\\
T^{26}&=&(R_1 (2 - f (2 - f (1 - R_1^2))))/(1 - f^2 R_1^2),\nonumber\\
 T^{65}&=&1-f, \quad T^{62}=(1-f)^2R_1/(1-f^2R_1^2),\qquad T^{63}=T^{64}=0.\nonumber\\
\end{eqnarray}
Replacing $R_1$ with $R_4$ in the above equation rest of the transmission probabilities $T^{31}$ to $T^{56}$ can be deduced.
\begin{figure}
  \centering    { \includegraphics[width=.94\textwidth]{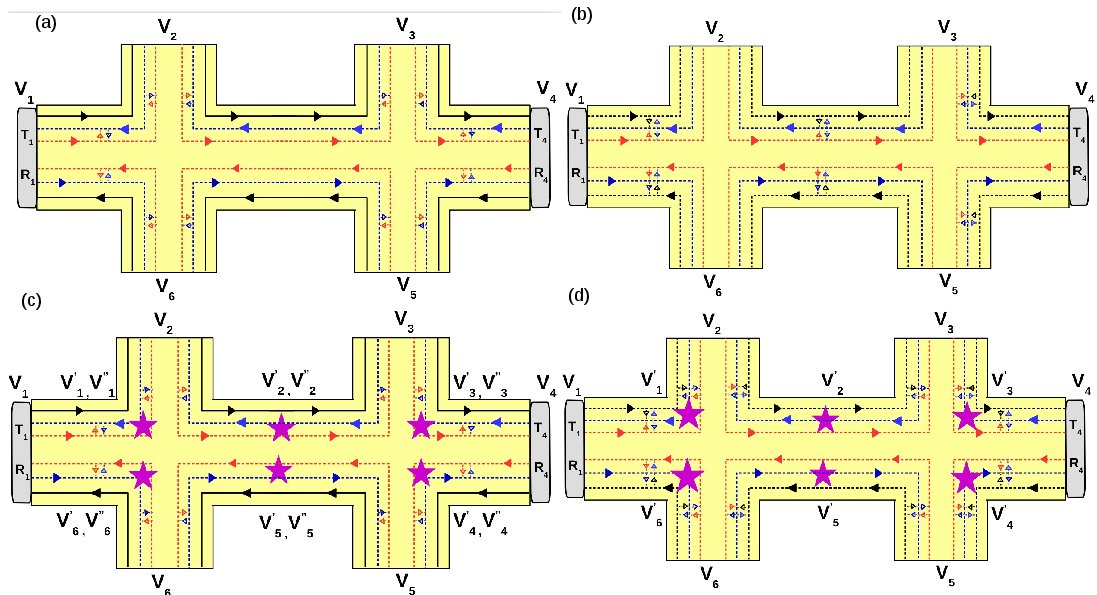}}\caption{ 6 terminal topological and trivial QAH edge mode along-with quasi-helical QSH edge modes. (a) Topological QAH edge mode with disorder at probes $1$ and $4$, (b) Trivial QAH edge mode with disorder at probes $1$ and $4$, (c) Topological QAH edge mode with inelastic scattering and disorder at probes $1$ and $4$, (d)  Trivial QAH edge mdoe with inelastic scattering and disorder at probes $1$ and $4$.}
\end{figure} 
Choosing reference potential $V_{4}=0$, and $I_2=I_3=I_5=I_6=0$ (as these are voltage probes), we derive longitudinal resistance $R_L^{Topo}=\frac{h}{e^2}\frac{2-3f+f^2}{9-15f+9f^2-2f^3}$ (for zero disorder). For finite disorder, the expression for $R^{Topo}_{L}$ is quite large, so we plot it in Fig.~8(a). 
\\

\subsubsection{ Effect of disorder  and inelastic scattering}
Herein we consider the effect of disorder and inelastic scattering on the various resistances for the sample as  shown in Fig.~7(c). The contacts $1$ and $4$ are disordered as in the previous case. The currents and voltages at the contacts from 1 to 6 are related by the equations-
\begin{eqnarray}
I_1&=&(3T_1-2T_1^2f/A)V_1-T_1V_6-T_1(1-f)/A(V'_1+V'_6),\nonumber\\
I_4&=&(3T_4-2T_4^2f/C)V_4-T_4V_3-T_4(1-f)/C(V'_4+V'_3),\nonumber\\
I_2&=&(3-2f)V_2-(1-f)(V'_1+V'_2)-V''_1,\nonumber\\
I_3&=&(3-2f)V_3-(1-f)(V'_3+V'_2)-V_2,\nonumber\\
I_5&=&(3-2f)V_5-(1-f)(V'_4+V'_5)-V''_4,\nonumber\\
I_6&=&(3-2f)V_6-(1-f)(V'_5+V'_6)-V_5,\nonumber\\
\end{eqnarray}
with $A=1-R_1f$ and $C=1-R_4f$, where the potential $V''_i$ are related to $V_i$ by-
\begin{eqnarray}
V''_2=V_2, \quad V''_5=V_5, \quad R_1V_6+T_1V_1=V''_1\nonumber\\
V''_3=V_3, \quad V''_6=V_6, \quad R_4V''_3+T_4V_4=V''_4
\end{eqnarray}
and the relation between potentials $V'_{i}$ and $V_{i}$ are mentioned in the Appendix.

Choosing reference potential $V_{4}=0$, and since 2,3, 5, 6 are voltage probes, $I_2=I_3=I_5=I_6=0$ we derive longitudinal resistance $R_L=\frac{h}{e^2}\frac{3-4f+f^2}{14-15f+6f^2-f^3}$ for zero disorder but finite inelastic scattering. The expression for $R_L$ in presence of both disorder and inelastic scattering is large so we analyze them via plots as in Fig.~8(b).

\subsection{Chiral(trivial) QAH edge mode with trivial QSH edge modes}

\subsubsection{ Effect of  disorder}
Herein again we consider two of the contacts $1$ and $4$ to be disordered, see Fig.~7(b). The current voltage relations are derived from the conductance matrix below:
\begin{equation}
G_{ij} =\frac{e^{2} }{h} \left( \begin{array}{cccccc}
    T^{11}  & -T^{12} & -T^{13} & -T^{14}  & -T^{15} & -T^{16}\\
    -T^{21}  & T^{22} & -T^{23} & -T^{24}& -T^{25} & -T^{26} \\ 
    -T^{31}  & -T^{32} & T^{33} & -T^{34}& -T^{35} & -T^{36} \\
     -T^{41}  & -T^{42} & -T^{43} & T^{44} & -T^{45} & -T^{46}\\ 
     -T^{51}  & -T^{52} & -T^{53} & -T^{54}& T^{55} & -T^{56} \\
     -T^{61}  & -T^{62} & -T^{63} & -T^{64} & -T^{65} & T^{66}\\ 
     \end{array} \right)
\end{equation}
where,
\begin{eqnarray}
 T^{11}&=&T_1 (3 - 2 (f + f_0) T_1/a - 2 (f^2 + f_0^2 + f f_0) T_1 R_1/a),\nonumber\\
T^{12}&=&((1 - f - f_0) T_1/a + (1 - f - f_0) R_1 T_1 (f + f_0)/a),\nonumber\\
T^{16}&=&(T_1 (2 - f - f_0) + a_1 T_1 R_1/a + a_1 T_1 R_1^2 (f + f_0)/a),\nonumber\\
T^{13}&=&T^{14}=T^{15}=0,\nonumber\\
T^{22}&=&(3 - 2 (f + f_0) - (1 - f - f0) R_1^2 a_1/a),\nonumber\\
T^{23}&=&(1 - f - f_0),\nonumber\\
T^{21}&=&(T_1 (2 - f - f_0) + T_1 R_1^2 (f + f_0) a_1/a + T_1 R_1 a_1/a),\nonumber\\
T^{26}&=&(R_1 ((1 - f)^2 + (1 - f_0)^2) + R_1^3 a_1^2/a),\nonumber\\
T^{25}&=&T^{24}=0,
\end{eqnarray}
with $a = 1 - R_2^2 (f^2 + f_0^2), c = 1 - R_4^2 (f^2 + f_0^2),  a_1 =  f (1 - f) + f_0 (1 - f_0)$. Replacing $R_1$ with $R_4$ in the above equation rest of the transmission probabilities $T^{31}$ to $T^{66}$ can be deduced. 
 
Choosing reference potential $V_{4}=0$, and $I_2= I_4=I_5=I_6=0$ (as $5,2,3$ and $6$ are voltage probes) we derive longitudinal resistance  $R_{L}^{Triv}=R_{23,14}=-((2 + f^2 - 3 f_0 + f_0^2 + f (-3 + 2 f_0))/(-9 + 2 f^3 + 15 f_0 -9 f_0^2 + 2 f_0^3 + f^2 (-9 + 6 f_0) + 3 f (5 - 6 f_0 + 2 f_0^2)))$ for ideal (zero disorder) case.  However, in general the expressions for $R_L$ are too large to reproduce here, so we will analyze them via plots as in Fig.~8(a).
\begin{figure}
 \centering  { \includegraphics[width=.6\textwidth]{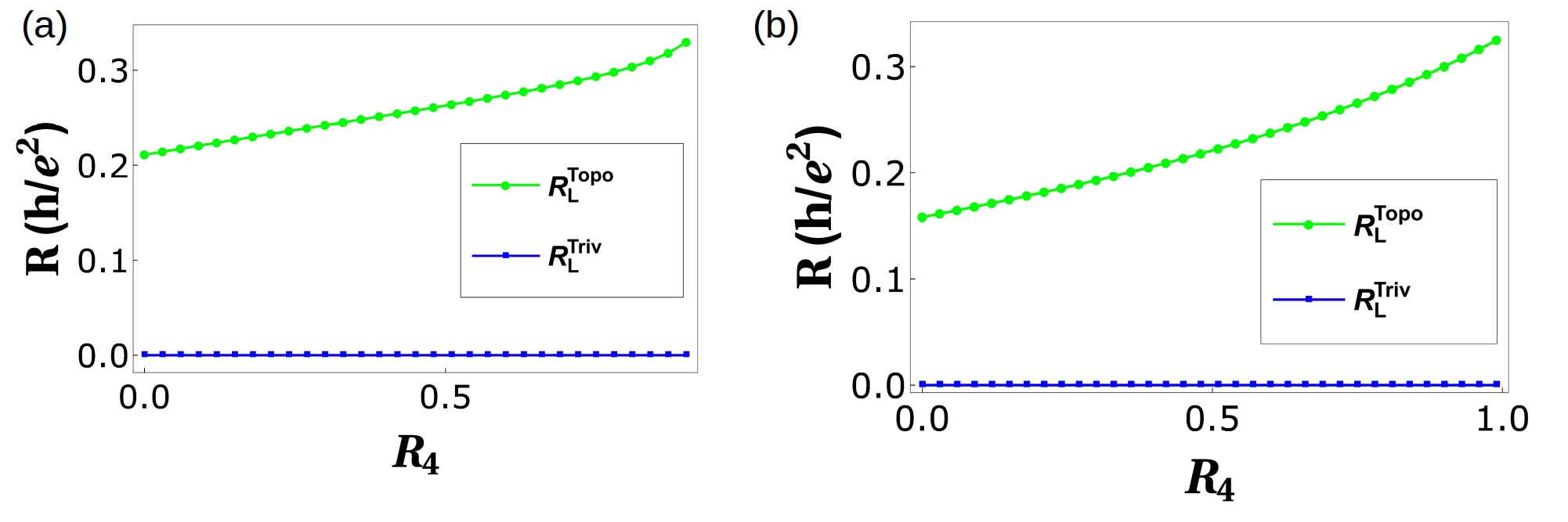}}
 \caption{(a) Longitudinal resistance $R_L$ vs. Disorder $R_{4}$ for parameters $R_1=0.5$ and spin-flip probability $f=f_0=0.5$, (b)  Longitudinal resistance  vs. Disorder $R_{4}$ in presence of inelastic scattering for parameters $R_1=0.5$ and spin-flip probability $f=f_0=0.5$. Note the longitudinal resistance vanishes for the trivial case but not for the topological case.}
\end{figure} 

\subsubsection{ Effect of disorder and inelastic scattering}
Herein we consider the trivial QAH edge modes with disorder and inelastic scattering, as shown in Fig.~7(d). Here the QAH chiral as well as the QSH helical edge modes are in the trivial phase, i.e. they can scatter from one edge mode to the other. All the edge modes interact with each other and via inelastic scattering equilibrate their energy to a  potential $V'_i$ ($`i'$ is from 1 to 6). 
The contacts $1$ and $4$ are disordered as in the previous case. The currents and voltages at the contacts from 1 to 6 are related by the equations-
{
\begin{eqnarray}
I_1&=&(3T_1-2T_1^2(f+f_0)/a-T_1^2R_1(f+f_0)^2/a-T_1^2R_1(f^2+f_0^2)/a)V_1\nonumber\\
&-&(T_1(1-f-f_0)/a+T_1R_1(1-f -f_0)(f + f_0)/a)V'_1-(T_1(2-f\nonumber\\
&-&f_0)+ T_1R_1(f (1 - f)+f_0(1-f_0))/a+T_1R_1^2(f(1 - f)\nonumber\\
&+&f_0(1-f_0))(f+f_0)/a)V'_6,\nonumber\\
I_2&=&(3 - 2 (f + f_0))V_2-(2 - (f + f_0))V'_1-(1 - (f + f_0))V'_2,\nonumber\\
I_3&=&(3 - 2 (f + f_0))V_3-(2 - (f + f_0))V'_2-(1 - (f + f_0))V'_3,\nonumber\\
I_5&=&(3 - 2 (f + f_0))V_5-(2 - (f + f_0))V'_4-(1 - (f + f_0))V'_5,\nonumber\\
I_6&=&(3 - 2 (f + f_0))V_6-(2 - (f + f_0))V'_5-(1 - (f + f_0))V'_6.\nonumber\\
\end{eqnarray}
The relation between potentials $V'_i$'s and $V_i$ are given in the appendix.
Choosing reference potential $V_{4}=0$, and as before $I_2=I_3=I_5=I_6=0$ (these are voltage probes), we derive the longitudinal resistance  $R_{L}^{Triv}=R_{23,14}$ in presence of inelastic scattering but for zero disorder as shown below-

\begin{eqnarray} 
R_{L}^{Triv}&=&-(((-3 + 2 f + 2 f_0) (2 + f^2 - 3 f_0 + f_0^2 + f (-3 + 2 f_0))^2)/(65 + 
     2 f^6 - 198 f_0 + 255 f_0^2 - 180 f_0^3 + 75 f_0^4 - 18 f_0^5 + 
     2 f_0^6 \nonumber\\&+& 6 f^5 (-3 + 2 f_0) + 15 f^4 (5 - 6 f_0 + 2 f_0^2) + 
     20 f^3 (-9 + 15 f_0 - 9 f_0^2 + 2 f_0^3) + 
     15 f^2 (17 - 36 f_0 + 30 f_0^2 - 12 f_0^3 + 2 f_0^4) \nonumber\\&+& 
     6 f (-33 + 85 f_0 - 90 f_0^2 + 50 f_0^3 - 15 f_0^4 + 2 f_0^5))).
\end{eqnarray}

The expression for longitudinal resistance in the general case of arbitrary disorder are quite large so we  examine it via plots as in Fig.~8(b). One thing is quite clear from Fig.~8(b), the case of trivial QAH edge mode with QSH quasi-helical goes over to single chiral(topological) QAH edge mode rather than the  topological QAH edge mode with QSH quasi-helical edge modes. This behavior replicated in the four terminal case too calls for a reinterpretation of the experimental results\cite{kou,che,wang}. In Table II we tabulate the results obtained for various cases for the longitudinal resistance. We also focus on the change due to change in magnetization from $\uparrow$ to $\downarrow$. There is a symmetry $R_{L}(\uparrow)=R_{L}(\downarrow)$ for trivial QAH edge mode which does not hold for a topological QAH edge mode with quasi helical QSH edge modes. This response of the trivial QAH edge mode is again in line with what was experimentally seen.
 {
\begin{center}
\begin{table}
\caption{Comparison of chiral(topological) QAH edge mode, chiral(topological) QAH edge mode with trivial  QSH edge modes and chiral(trivial) QAH edge mode with trivial  QSH edge modes, see also Fig.~8(a,b).}
\begin{tabular}{ |c|c|c|c|p{3.8cm}|}
\hline
&&QAH(topological)&QAH(topological)+QSH(trivial)&QAH(trivial)+QSH(trivial)  \\ 
 \hline
Zero disorder&$R_L$&0 ($R_{L}(\uparrow)=R_{L}(\downarrow)$)&Finite ($R_{L}(\uparrow)=R_{L}(\downarrow)$) & Finite ($R_{L}(\uparrow)=R_{L}(\downarrow)$)\\
\hline
Disordered probes &$R_L$&0 ($R_{L}(\uparrow)=R_{L}(\downarrow)$)&Finite ($R_{L}(\uparrow)=R_{L}(\downarrow)$)&0 ($R_{L}(\uparrow)=R_{L}(\downarrow)$)\\
\hline
Disorder+inelastic scattering &$R_L$&0 ($R_{L}(\uparrow)=R_{L}(\downarrow)$) &Finite ($R_{L}(\uparrow) \neq R_{L}(\downarrow)$)& 0 ($R_{L}(\uparrow)=R_{L}(\downarrow)$)\\
\hline
\end{tabular}
\end{table}
\end{center}
}
 
\section{Conclusion} 
We conclude by analyzing the tables I and II. We see that the trivial(chiral) QAH edge mode with trivial quasi-helical QSH edge modes is more closer to the experimental situation, as interpreted in Ref.~\cite{wang} than the topological(chiral) QAH edge mode with trivial quasi-helical QSH edge modes is. This implies a reevaluation of the consensus regarding those quantum anomalous Hall experiments\cite{kou,che,wang1}. Perhaps, something else is happening and maybe these are not true chiral(topological) quantum anomalous Hall edge modes which were seen, but rather what could be described as chiral(trivial) QAH edge modes.


\section{Acknowledgments}
This work was supported by funds from Dept. of Science and Technology (SERB), Govt. of India, Grant No. EMR/2015/001836.

\section*{Author contributions statement}
C.B. conceived the proposal,  A.M. and C.B. wrote the paper, A.M. and C.B analyzed the results. 

\section{ Appendix(Supplementary Material): Relation between equilibrated potentials and probe potentials}
 \subsection{ Relation between $V_i '$ and $V_i$ for Four terminal QAH bar with chiral(trivial) QAH edge mode and trivial QSH edge modes}
 See Section 2.3.2 and Fig.~4(b) of the manuscript for the details of the set-up. See Eq.~16 of the manuscript for relations between currents and voltages.
 \begin{eqnarray}
&(2 - f - f_0)V_1+\frac{[1 -( f + f_0)] [1 + (f + f_0) R_2] T_2}{[1-R_2^2(f^2+f_0^2)]}V_2+\frac{R_2(1-f-f_0)^2}{[1-R_2^2(f^2+f_0^2)]}V'_2=[(2 - f - f_0)+\frac{[1 -( f + f_0)] [1 + (f + f_0) R_2] T_2}{[1-R_2^2(f^2+f_0^2)]}+\frac{R_2(1-f-f_0)^2}{[1-R_2^2(f^2+f_0^2)]}]V'_1,\nonumber\\
&\frac{[2 - (f + f_0) + ((1 - f) f + (1 - f_0) f_0) R_2 - (f - f_0)^2 R_2^2) T_2}{[1-R_2^2(f^2+f_0^2)]}V_2+[1-(f+f_0)]V_3+\frac{(2 + (-2 + f) f + (-2 + f_0) f_0) R_2 + (f - f_0)^2 R_2^3}{[1-R_2^2(f^2+f_0^2)]}V'_1=\nonumber\\&[\frac{[2 - (f + f_0) + ((1 - f) f + (1 - f_0) f_0) R_2 - (f - f_0)^2 R_2^2) T_2}{[1-R_2^2(f^2+f_0^2)]}+[1-(f+f_0)]+\frac{(-2+f)f +(-2 +f_0)f_0)R_2 + (f - f_0)^2 R_2^3}{[1-R_2^2(f^2+f_0^2)]}]V'_2,\nonumber\\
&(2 - f - f_0)V_3+\frac{[1 -( f + f_0)] [1 + (f + f_0) R_4] T_4}{[1-R_4^2(f^2+f_0^2)]}V_4+\frac{R_2(1-f-f_0)^2}{[1-R_2^2(f^2+f_0^2)]}V'_4=[(2 - f - f_0)+\frac{[1 -( f + f_0)] [1 + (f + f_0) R_2] T_2}{[1-R_2^2(f^2+f_0^2)]}+\frac{R_2(1-f-f_0)^2}{[1-R_2^2(f^2+f_0^2)]}]V'_3,\nonumber\\
&\frac{[2 - (f + f_0) + ((1 - f) f + (1 - f_0) f_0) R_4- (f - f_0)^2 R_4^2) T_4}{[1-R_4^2(f^2+f_0^2)]}V_4+[1-(f+f_0)]V_1+\frac{(2 + (-2 + f) f + (-2 + f_0) f_0) R_4 + (f - f_0)^2 R_4^3}{[1-R_4^2(f^2+f_0^2)]}V'_3=\nonumber\\&[\frac{[2 - (f + f_0) + ((1 - f) f + (1 - f_0) f_0) R_4 - (f - f_0)^2 R_4^2) T_4}{[1-R_4^2(f^2+f_0^2)]}+[1-(f+f_0)]+\frac{(-2+f)f +(-2 +f_0)f_0)R_4 + (f - f_0)^2 R_4^3}{[1-R_2^2(f^2+f_0^2)]}]V'_4.\nonumber\\
\end{eqnarray}

\subsection{ Relation between $V_i '$ and $V_i$ for six terminal QAH bar with chiral(topological) QAH edge mode and trivial QSH edge modes}
 See Section 3.2.2 and Fig.~7(c) of the manuscript for the details of the set-up. See Eq.~22 of the manuscript for relations between currents and voltages.
\begin{eqnarray}
&(1-f)(V_2+V_3)=2(1-f)V'_2,\quad (1-f)(V_5+V_6)=2(1-f)V'_5,\nonumber\\
&(1-f)V_3+(\frac{T_4(1-f)}{(1-R_4^2f^2)}+\frac{T_4R_4f(1-f)}{1-R_4^2f^2})V_4+\frac{R_4(1-f)^2}{(1-R_4^2f^2)}V'_4=((1-f)+(\frac{T_4(1-f)}{(1-R_4^2f^2)}+\frac{T_4R_4f(1-f)}{1-R_4^2f^2})+\frac{R_4(1-f)^2}{(1-R_4^2f^2)})V'_3,\nonumber\\
&(1-f)V_6+(\frac{T_1(1-f)}{(1-R_1^2f^2)}+\frac{T_1R_1f(1-f)}{1-R_1^2f^2})V_1+\frac{R_1(1-f)^2}{(1-R_1^2f^2)}V'_1=((1-f)+(\frac{T_1(1-f)}{(1-R_1^2f^2)}+\frac{T_1R_1f(1-f)}{1-R_1^2f^2})+\frac{R_1(1-f)^2}{(1-R_1^2f^2)})V'_3,\nonumber\\
&(1-f)V_5+(\frac{T_4(1-f)}{(1-R_4^2f^2)}+\frac{T_4R_4f(1-f)}{1-R_4^2f^2})V_4+\frac{R_4(1-f)^2}{(1-R_4^2f^2)}V'_3=((1-f)+(\frac{T_4(1-f)}{(1-R_4^2f^2)}+\frac{T_4R_4f(1-f)}{1-R_4^2f^2})+\frac{R_4(1-f)^2}{(1-R_4^2f^2)})V'_3,\nonumber\\
&(1-f)V_2+(\frac{T_1(1-f)}{(1-R_1^2f^2)}+\frac{T_1R_1f(1-f)}{1-R_1^2f^2})V_1+\frac{R_1(1-f)^2}{(1-R_1^2f^2)}V'_6=((1-f)+(\frac{T_1(1-f)}{(1-R_1^2f^2)}+\frac{T_1R_4f(1-f)}{1-R_1^2f^2})+\frac{R_1(1-f)^2}{(1-R_1^2f^2)})V'_3.\nonumber\\
\end{eqnarray}
with $a = 1 - R_2^2 f^2, c = 1 - R_4^2 f^2$.
\subsection{ Relation between $V_i '$ and $V_i$ for six terminal QAH bar with chiral(trivial) QAH edge mode and trivial QSH edge modes}
See Section 3.3.2 and Fig.~7(d) of the manuscript for the details of the set-up. See Eq.~26 of the manuscript for relations between currents and voltages.

\begin{eqnarray}
(T_1(2-f-f_0)+T_1R_1(f(1-f)&+&f_0(1-f_0))/a +T_1R_1^2(f + f_0)(f(1-f)+f_0(1-f_0))/a)V_1+(1-f-f_0)V_2\nonumber\\+(R_1((1-f)^2+(1-f_0)^2)&+&R_1^3((1-f)f+f_0(1-f_0))^2/a)V'_6= (T_1(2-f-f_0)+T_1R_1(f(1-f)+f_0(1-f_0))/a \nonumber\\+T_1R_1^2(f+f0)(f(1-f)+f_0(1-f_0))/a&+&(1-f-f_0)+R_1((1-f)^2+(1-f_0)^2)+R_1^3((1-f)f+f_0(1-f_0))^2/a)V'_1,\nonumber\\
(2-f-f_0)V_6+(T_1(1-f-f_0)/a&+&T_1R_1(f+f_0)(1-f-f_0)/a)V_1+R_1(1-f-f_0)^2/aV'_1=((2-f-f_0)\nonumber\\+(T_1(1-f-f_0)/a+T_1R_1(f+f_0)&&(1-f-f_0)/a)+R_1(1-f-f_0)^2/a )V'_6,\nonumber\\
(2-f-f_0)V_2+(1-f-f_0)V_3&=&(3-2(f+f_0))V'_2,\nonumber\\
(2-f-f_0)V_5+(1-f-f_0)V_6&=&(3-2(f+f_0))V'_5,\nonumber\\
(T4(2-f-f_0)+T_4R_4(f(1-f)&+&f_0(1-f_0))/c+T_4R_4^2(f+f_0)(f(1-f)+f_0(1-f_0))/c)V_4+(1-f-f_0)V_5\nonumber\\+(R_4((1-f)^2+(1-f_0)^2)&+&R_4^3((1-f)f+f_0(1-f_0))^2/c)V'_3=(T_4(2-f-f_0)+T_4R_4 (f (1 - f) + f_0 (1 - f_0))/c \nonumber\\+T_4 R_4^2 (f + f_0) (f (1 - f) &+& f_0 (1 - f_0))/c + (1 - f - f_0) +R_4 ((1 - f)^2 + (1 - f_0)^2) +R_4^3 ((1 - f) f + f_0 (1 - f_0))^2/c) V'_4,\nonumber\\
(2 - f - f_0) V3 + (T_4 (1 - f - f_0)/c& +&T_4 R_4 (f + f_0) (1 - f - f_0)/c) V_4 +R_4 (1 - f - f_0)^2/c V'_4 = ((2 - f - f_0) \nonumber\\+ (T_4 (1 - f - f_0)/c +T_4 R_4 (f + f_0)&&(1 - f - f_0)/c) +R_4 (1 - f - f_0)^2/c ) V'_3
\end{eqnarray}

 \textbf{Competing financial interests} 

The author(s) declare no competing financial interests.

\textbf{Data availability statement}
As this is an analytical work, all the data needed to plot the Figs. 3,5, 6 and 8 are generated from the expressions shown in the manuscript.

\end{document}